# A spiral laser scanning routine for powder bed fusion inspired by natural predator-prey behavior

**Suh In Kim\* and A. John Hart\***

Department of Mechanical Engineering, Massachusetts Institute of Technology, 77 Massachusetts Avenue, Cambridge, MA, 02139 USA

## Abstract

Additive manufacturing by laser powder bed fusion (LPBF) offers material versatility, capability for complex geometries, and control of mechanical properties and microstructure. However, achieving high-quality output requires process parameters that consider both local and global thermal gradients. Here, we propose a new scan pattern for mitigating part quality issues caused by non-uniform heating and cooling. A nature-inspired design method is employed to derive a spiral pattern stemming from the predator-prey behavior, and a power optimization routine is applied to maintain constant melt pool depth. Comparing simulated thermal histories for the spiral pattern to well-established zig-zag and helix scan patterns, we propose that the spiral pattern significantly reduces the spatial variation of temperature across the scan area, while a larger area remains above a specified threshold temperature at the end of the scan. Consequently, the spiral pattern is promising for LPBF of crack-prone materials, and for future optimization of LPBF overall.

\* Corresponding authors.
  *E-mail address:* sikim@mit.edu (Suh In Kim), ajhart@mit.edu (A. John Hart)

# 1. Introduction

The laser powder bed fusion (LPBF) additive manufacturing (AM) process is increasingly adopted due to its capability for highly complex geometries and its applicability to a wide variety of metals (Gisario et al. 2019). However, key challenges of implementing LPBF arise from the local and global thermal history imparted by the laser scan pattern, and requisite management of residual stresses (Hussein et al. 2013), mitigation of defects and microcracks (Han et al., 2018), and attainment of spatially homogeneous microstructure (Dezfoli et al. 2017). Generally, spatially uneven heating and cooling can result in microcracking caused by severe residual stresses; this is particularly problematic in alloys with high melting temperature, high thermal conductivity, and/or high DBTT (ductile-to-brittle transition temperature) such as superalloys and refractory alloys (Vrancken et al. 2020). More generally, even in cases where low porosity, geometrically accurate components are formed by LPBF, heat treatment is employed to relieve residual stress and to homogenize and/or alter the microstructure according to the design intent (Qiu et al. 2013; Han et al., 2018; Roehling et al. 2019). Mitigating thermal inhomogeneity in the printing process can increase the number of applicable high-temperature materials in LPBF, improve dimensional accuracy, and reduce required post heat treatment efforts.

LPBF traditionally employs the "island" scanning strategy, which was originally shown by Kruth et al. to prevent the high residual stresses caused by long, continuous scan vectors (Kruth et al. 2004). Generally, each layer is divided into rectangles, each filled by a zig-zag pattern. The scanning directions usually change 90 degrees in adjacent islands and the entire pattern is rotated 67 degrees in subsequent layers. The island scanning strategy is known as effective in mitigating the anisotropy and accumulation of residual stresses. Yet, because each track is cyclically heated, the concentration of heat at the reversal points of the zig-zag, and in border areas of the part can lead to overheating, deformation, and porosity. As a result, others have proposed alternative scan patterns such as a space filling method with Hilbert or Gosper patterns (Yu et al. 2011; Babu et al. 2015; Catchpole-Smith et al. 2017). The space filling method was shown to be effective in reducing the residual stresses, keeping a consistent melt pool geometry, and mitigating the micro-cracking problems found in non-weldable materials (Catchpole-Smith et al. 2017). Besides the pattern design, the scanning direction and length can be critical factors affecting printing quality, particularly in dealing with the geometries prone to overheating (Parry et al. 2019). The selection of local laser power and scan speed also highly affect the local surface qualities (Mumtaz and Hopkinson 2010; Yeung et al. 2019) and melt pool stability (Martin et al. 2019). Thus, the scan pattern design, followed by the optimization of process parameters, is required to manage thermal history and give desired geometric and material quality in LPBF.

Nevertheless, to our knowledge all presently realized scan patterns have been derived from intuition and iteration rather than algorithmic approaches. Bossier et al presented a model-based control approach to derive a conceptual path for the printing region of arbitrary shape, which is not suitable for the island strategy (Boissier et al. 2020). A scan optimization approach based on a genetic algorithm was also developed for the electron beam melting (EBM) process, but it only considered a spot scanning strategy (Halsey et al. 2020), which is impractical for the LPBF process that prefers continuous scan lines due to the need to position the laser with moving mirrors. Additionally, scan strategies for LPBF should be applicable to machines with multiple coordinated yet independently scanned lasers (Masoomi et al. 2017; Zhang et al. 2020).

The main challenge of optimizing the scan pattern for LPBF is the complexity of the optimization problem, which is similar to a Traveling Salesman Problem (TSP). Classically, the TSP concerns the optimal visiting sequence of cities (Dorigo and Gambardella 1997), which is analogous to a visiting sequence for the grid points composing a scan pattern. In addition to the conventional TSP problem, the scan pattern design problem for LPBF needs to consider a jump motion between scan vectors, which corresponds to the movement of the mirrors or other mechanism that positions the laser, while the laser power is off. Accordingly, the number of candidate paths when designing for LPBF scanning is much larger than conventional TSP problems. To estimate the performance of each candidate scan pattern, a computationally intensive thermal analysis is required. Therefore, finding an optimal scan pattern for the LPBF process by a conventional optimization approach is prohibitive.

In this paper, we propose an alternative scan pattern for LPBF, inspired by the adaptation of a predator-prey scheme to the goal of optimizing the local thermal history of the LPBF process. An analogy is made between predation in natural systems and the dynamics of laser heating and subsequent cooling, and then the result of the adapted predator-prey solution is used to guide the definition of a rotating spiral unit cell pattern for use in LPBF. From this, scan parameters are defined and a thermal finite element model is applied to simulate and derive an optimization scheme for laser power enabling the consistent melt pool depth. Then, the performance of the rotating spiral pattern is compared to established scan patterns based on simulation results.

## 2. Synthesis of a scan pattern for LPBF inspired by predator-prey behavior

This study began by noticing the qualitative similarity between the cyclic interaction of predator and prey populations in ecological systems, and the cyclic thermal conditions that arise due to scan-wise and layer-wise heating and cooling in LPBF. Ecological systems exhibit persistent cyclic motions, as a resilience to stochastic events (Blasius et al. 2020). The well-known predator-prey models (Lotka, 1925; Volterra, 1926; Gause et al. 1936) mathematically represent the time-varying populations, as schematically shown in Figure 1(a). Meanwhile, Figure 1(b) represents the time history of the LPBF process for a fixed observation point along a scan vector (Yang et al. 2018). The temperature at the observation point rapidly increases after the heat source passes though the point, with a slight time delay. This time history is similar to that of the predator-prey model depicted in Figure 1(a), where the prey and predator can be respectively regarded as the laser power and temperature. Likewise, there is a similarity in terms of the spatial dynamics in the two systems. The spatial predator-prey model includes diffusion terms for the predator and prey, where the diffusion model for the predator is similar to the thermal diffusion in the LPBF process. Last, the spatial predator-prey model can also include a competition model for the prey (Britton 1989). In the case of a non-local competition model, the population of prey living in a finite area can be limited not to exceed a specific value set by the model parameter. In the LPBF process, this limitation of the prey population can, for instance, be matched to a machine limitation for the total laser power.

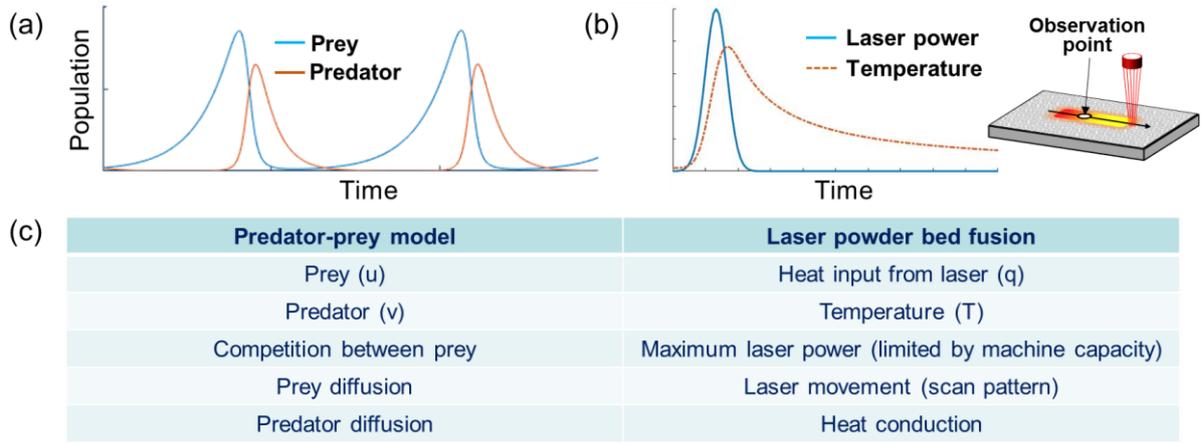

**Figure 1.** (a) Exemplary time history of the predator and prey population, and (b) time histories of power and temperature observed at a point on the scan vector. (c) Adaptation of variables in the predator-prey model to the LPBF process.

Broadly, a goal of toolpath design and optimization for LPBF is to maintain thermal uniformity in the presence of the spatially and temporally varying application of laser energy. Therefore, when one or more lasers are used, it is desirable to have a scan pattern that avoids local overheating, while maintaining a high throughput and ensuring that all points in the scan are melted. Given this strategy, one can apply the predator-prey analogy when seeking efficient management of the heat source, because the prey (laser) move in their habitat to avoid the predator (temperature). That is, an interaction between two species, included in the predator-prey model, could result in qualitative insights of how to achieve optimal laser control for LPBF.

Thus, we implemented a simulation of the predator-prey model involving non-local competition between prey to derive a conceptual scan pattern for the LPBF process. The spatiotemporal patterns of the prey and predators in Figure 2(a) were obtained by numerical simulation of equation (1), below.

$$u_{,t} = \alpha u \left(1 - \frac{1}{2u_0 A} \iint_{\Omega_\mathbf{x}} u \, d\mathbf{x}\right) - \beta u v + D_u (u_{,xx} + u_{,yy}) \quad (1a)$$

$$v_{,t} = -\gamma v + \delta u v + D_v (v_{,xx} + v_{,yy}) \quad (1b)$$

Here, $u(\mathbf{x}, t)$ and $v(\mathbf{x}, t)$ are populations of the prey and predator, respectively. In model (1a), the coefficient $\alpha$ controls the rate of prey's birth. Two coefficients, $\beta$ and $\delta$ of the bilinear terms in (1) determine the decrease/increase rate of population by predation. The coefficient $\gamma$ representings the rate of the predator's death; and $D_u$ and $D_v$ are the diffusivity of prey and predator, respectively. An interval of integration in (1a), i.e., $\Omega_\mathbf{x}$, corresponds to the boundary size of the prey's internal competition. The population of prey living in a habitat $\Omega_\mathbf{x}$ will not exceed $2u_0 A$ due to the competition model, where $A$ is the size of the competition boundary, and $u_0$ is a so-called equilibrium point defined by $\gamma/\delta$.

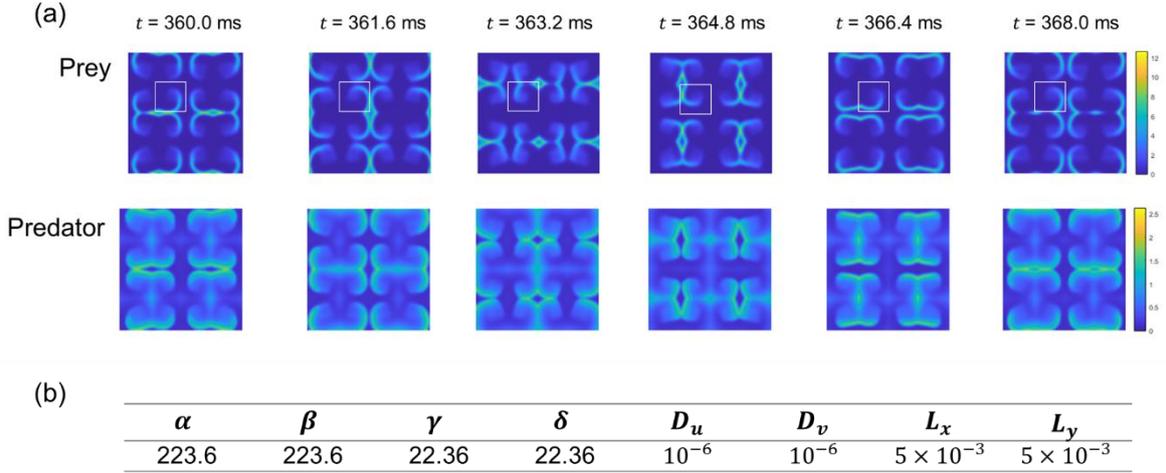

**Figure 2.** (a) A spatiotemporal pattern obtained from the simulation of a predator-prey model with the nonlocal competition effect. A local region in the white-colored box (plotted in the prey population map) shows a rotating spiral spatiotemporal pattern. (b) Material property values applied to the simulation of the adapted predator-prey model.

Equation (2) corresponds to the thermal diffusion equation addressing the LPBF process. Here, $k$ is the thermal conductivity, $\rho$ the density, $c_p$ the specific heat. Additionally, we define $\tilde{\alpha} = k/(\rho c_p)$, which is the thermal diffusivity corresponding to $D_v$ of (1b).

$$\rho c_p T_{,t} = k(T_{,xx} + T_{,yy} + T_{,zz}) + q \qquad (2)$$

Equations (1b) and (2) are similar, but the dimensions of both equations are different. The predator-prey model is two-dimensional, but heat conduction in LPBF is three-dimensional. Instead of the thermal diffusion in the Z-direction, the adapted predator model in (1b) has a decaying term represented by $-\gamma v$, which will be able to approximately simulate the Z-direction heat loss (toward the baseplate) in the LPBF process. Additionally, in (1b), $\delta uv$ can be regarded as a correspondence to a heat source $q$ in (2); when laser power (prey) is given at a point $\mathbf{x}$, the temperature will increase.

The thermal diffusivity of metals typically used in LPBF has a value between $10^{-5}$ and $10^{-6}$. Thus, we set the diffusivity of prey and predator to be in the same range. We also set $\gamma$ to similarly describe the temperature decaying speed. In the case of the competition boundary, $\Omega_{\mathbf{x}}$ is set to $\Omega_{\mathbf{x}} = \{\mathbf{y} | \|\mathbf{x} - \mathbf{y}\|_p \leq R\}$. Here, we applied $p = 1$ and $R = 1.75 \times 10^{-3}$. Figure 2(b) lists all parameter values for the present implementation of the model, where $L_x$ and $L_y$ are the domain size. Also, the simulation applied a forward Euler scheme with Neumann boundary condition corresponding to zero flux at the boundary.

The result of the predator-prey model simulation in Figure 2 shows a symmetric pattern composed of local spirals. The simulation starts with the initial prey and predators uniformly distributed over the domain except for the center point. For the excitation of the field variables, the initial value of the prey at the center point is set to five times of the equilibrium point, $u_0$. As a result, a heterogeneous pattern of both species emerges due to the diffusion occurring at the center point. For reference, the predator-prey model can result in a trivial solution for each population, which converges to an equilibrium point or zero value. For example, extinction occurs when the predator population converges to a zero-population solution. On the other hand, a previous study has shown that diffusion terms applied to the predatory-prey model bring on an oscillation called chaotic motion (Pascual 1993). Additionally, several studies have discussed that a certain set of model parameters can result in a heterogeneous or an oscillating solution based on the bifurcation theory (Turing 1990; Baurmann et al. 2007).

The model parameters that we applied to the LPBF adaptation also showed the oscillating heterogeneous solution, in Figure 2. The spiral pattern is a result of the interaction between the prey and predator, moving in the given habitat (Gurney et al. 1998; Hawick et al. 2008). Like a dog that cannot catch its tail despite endless efforts with rotations, this spiral-like pattern can provide a sustainable balance between the prey and predator without termination (i.e., the cyclic fluctuation of populations without distinction of both species).

The predator-prey model outputs a spatial pattern which would require imaging of energy input over the considered area, so now we adapt the spiral pattern to be compatible with raster-scanning or a point input as used in LPBF. This is shown in Figure 3(a,b). Previous studies have proposed helix scan patterns for LPBF (Bo et al. 2012; Cheng et al. 2016; Hagedorn-Hansen et al. 2017); however, these gradually approach the inside (or outside) of the scan region, while our pattern allows each scan vector to traverse from the perimeter to the center point. After each vector ends, the laser jumps to the starting point of the next vector, repeating the out-in sequence. Also, Ramos et al. (Ramos et al. 2019) proposed a parallel line segment pattern where the order of scanning alternated from the outside-in. Figure 3(c) illustrates the reference scan patterns.

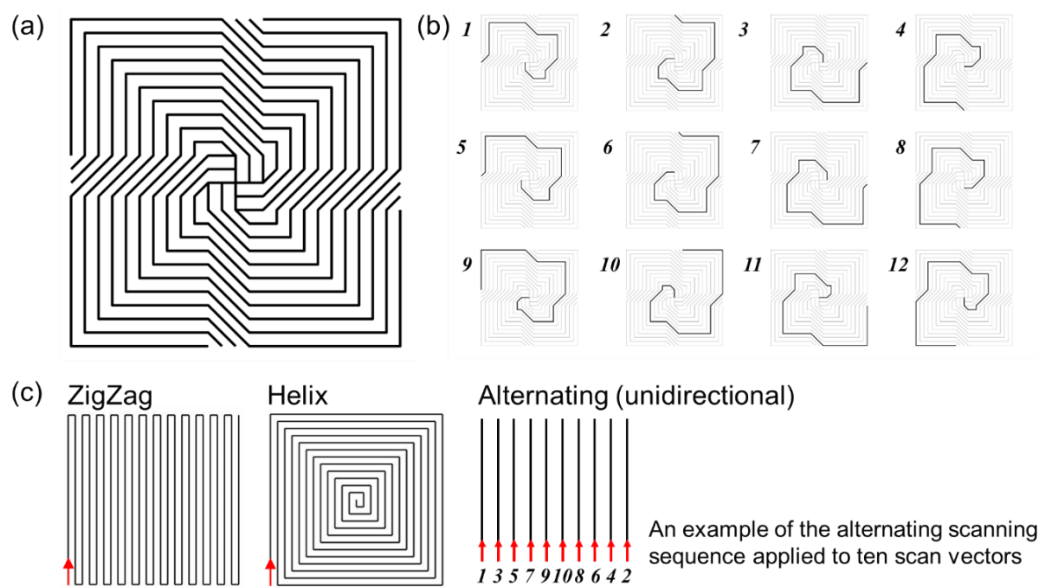

**Figure 3.** (a) A rotating spiral scan pattern inspired by the spatiotemporal pattern of the prey population appearing in the predator-prey simulation. (b) A sequence of twelve scan vectors composing the rotating spiral scan pattern. A laser draws the scan pattern sequentially, from outside to inside (or reversely). The laser will be turned off at the end of each scan vector and turned on again after moving to the start point of the next scan vector. (c) Two benchmark scan patterns, called zigzag (left) and helix (right) and an alternating unidirectional scan pattern introduced by Ramos et al. (Ramos et al. 2019).

Our proposed spiral scan pattern is incidentally similar to the design of frequency selective electromagnetic surfaces (Yin et al. 2019). As a scan pattern for LPBF, it is attractive because of the following reasons related to its geometrical features: (i) This pattern does not contain long, linear scan vectors, which could induce high-residual stresses (Kruth et al. 2004; Ramos et al. 2019). (ii) Unlike the previously designed helix patterns (Hagedorn-Hansen et al. 2017), each scan vector traverses from outside to inside (or reversely), and sequentially the vectors rotate to mimic the output of the predator-prey model in Figure 2. (iii) The pattern can be tuned by the number of scan vectors, i.e., to reduce the hatch distance the pattern can be preserved by increasing the number of vectors constituting the spiral unit cell.

While the true output of the predator-prey simulation (Figure 2) would require exposure of the entire area simultaneously, we suggest that the spiral pattern can mimic the result shown in Figure 2 even with a finite scan speed machine. Still, the instability inherent in the melt pool dynamics of the LPBF process limits the scan speed (Gunenthiram et al. 2017; Saunders 2017). Additionally, the scan area in Figure 3(a) is compact relative to the scan vector length to promote thermal uniformity also from the sequential rotation of the scan vectors.

## 3. Optimization of laser power based on a semi-analytical simulation

A well-designed scan pattern evenly distributing energy will reduce the temperature variance over the given scan area. However, even a low variance of temperature over the scan area does not ensure the uniform temperature history in the vicinity of the melt pool over the whole scanning process. With both traditional (zig-zag) scanning, and any alternative pattern, a non-uniform temperature distribution is detrimental to the quality output of LPBF. For instance, local hot spots at the hairpin ends of adjacent zig-zag scan vectors can cause excess spatter generation and/or generation of keyhole porosity (Martin et al. 2019; Khairallah et al. 2020).

Therefore, following derivation of the spiral pattern, we now present a method for optimization of the laser parameters (namely, power and scan speed) to achieve thermal uniformity and to avoid over/under-heating issues. Overheating by a high energy density may cause undesirable trapped pores (Martin et al. 2019). Additionally, a low energy density can cause pores due to lack of fusion (Mukherjee and DebRoy 2018). To prevent pore formation due to local under- or overheating, several laser power control techniques have been developed in the LBPF process. These techniques comprise model/rule-based schemes (Martin et al. 2019; Yeung et al. 2019; Yeung and Lane 2020; Druzgalski et al. 2020) and iterative correction methods based on simulation models. For this work, a simplified and fast simulation model is needed for the temperature distribution in the vicinity of the laser spot as scanning occurs (Peng et al. 2018; Yang et al. 2018; Yavari et al. 2019; Yang et al. 2020).

To establish scan parameters that maintain constant melt pool depth, we implement an optimization method with a semi-analytical approach, as described by Yang et al. (Yang et al. 2018; Yang et al. 2020). This incorporates a linear heat conduction model with constant material properties, as described in (2), and the linear assumption enables the semi-analytical method to use superposition for calculating the temperature field at time $t$, as shown in (3). With this superposition method, the temperature field can be calculated based on an analytical solution $\tilde{T}(\mathbf{x}, t)$ in (4) and a numerical solution $\hat{T}(\mathbf{x}, t)$ that will be calculated based on (5).

The model is more time-efficient than the fully numerical method requiring a fine mesh around the melt pool accompanying high temperature gradients. Even without the fine mesh in the vicinity of the melt pool, the semi-analytical method can accurately analyze the rapidly changing temperature field by employing the exact analytical solution near the melt pool in (4a).

$$T(\mathbf{x},t) = \tilde{T}(\mathbf{x},t) + \hat{T}(\mathbf{x},t) \tag{3}$$

$$\tilde{T}^{(I)}(\mathbf{x},t) = \frac{Q^{(I)}\tilde{A}}{4\rho c_p (\pi\tilde{\alpha}(t-\tau_0^{(I)}))^{3/2}} \exp\left(\frac{-(R^{(I)})^2}{4\tilde{\alpha}(t-\tau_0^{(I)})}\right) \tag{4a}$$

$$\tilde{T}(\mathbf{x},t) = \sum_{I=1}^{M} \tilde{T}^{(I)}(\mathbf{x},t) \quad for \ t > t_0^{(M)} \tag{4b}$$

$$\rho c_p \hat{T}_{,t} = k\left(\hat{T}_{,xx} + \hat{T}_{,yy} + \hat{T}_{,zz}\right) \quad \text{in V} \tag{5a}$$

$$\hat{T} = T_{base} - \tilde{T} \quad \text{on } \partial V_{bot} \tag{5b}$$

$$-k\nabla\hat{T} \cdot \mathbf{n} = k\nabla\tilde{T} \cdot \mathbf{n} \quad \text{on } \partial V_{lat} \tag{5c}$$

$$-k\nabla\hat{T} \cdot \mathbf{n} = 0 \quad \text{on } \partial V_{top} \tag{5d}$$

In (4), $\tilde{T}^{(I)}$ is the analytical solution for the temperature field at time $t$ induced by the incident laser spot heat source, $Q^{(I)}$, applied at point $\mathbf{x}^{(I)}$ on time $t_0^{(I)}$, where the distance $R^{(I)}$ is calculated by $\|\mathbf{x} - \mathbf{x}^{(I)}\|_2$ and $\tilde{A}$ is the absorptivity fraction of the laser. In (4a), $\tau_0^{(I)}$ is equal to $t_0^{(I)} - r^2/(8\tilde{\alpha})$, where $r$ is the radius of a laser spot. The total analytical solution $\tilde{T}$ is the summation of the solutions in (4a) for the heat sources applied before time $t$.

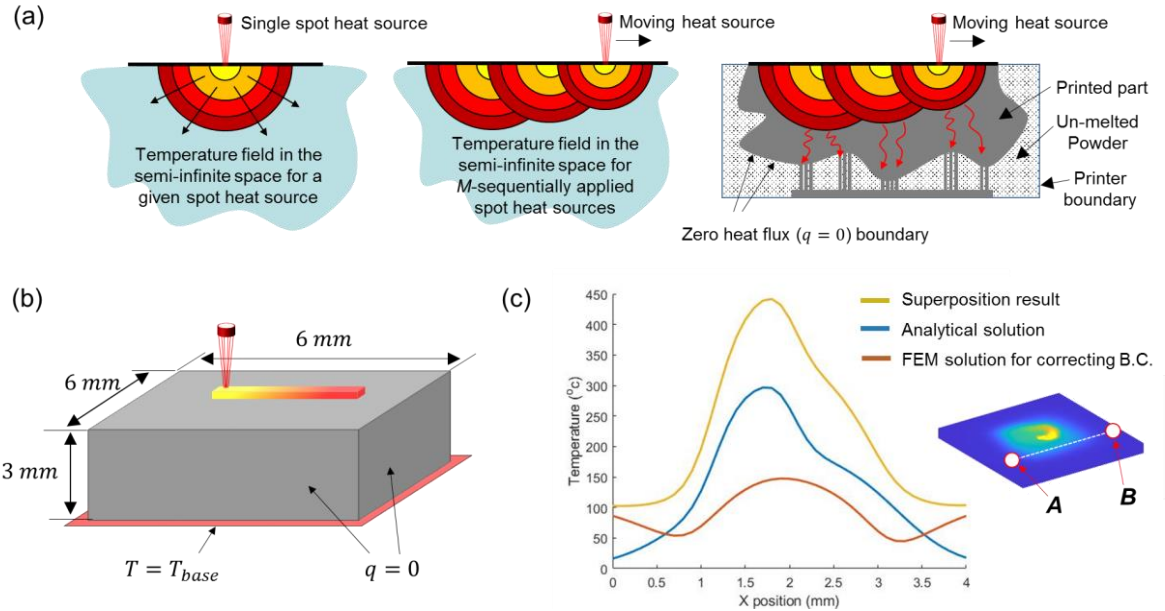

**Figure 4.** (a) Illustration of the basic concept of the semi-analytical analysis. A set of spot heat sources applied with a given time interval can simulate a moving heat source traversing the scanning area along the scan vector. (b) Boundary conditions applied to the simulation model, (c) the result of the semi-analytical simulation method. The plot corresponds to the temperature distributions along the line connecting the points A and B.

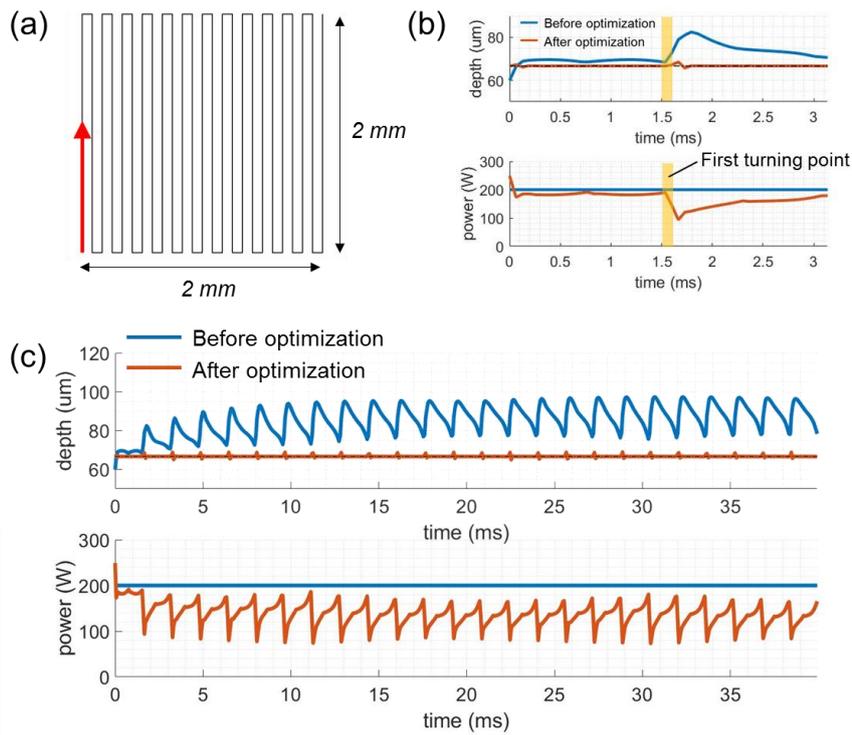

**Figure 5.** (a) A zig-zag scan pattern applied to power optimization. (b) The laser power and melt pool depth graphs before and after optimization for the first two scan vectors involving the first turning point, and (c) over the whole scanning process.

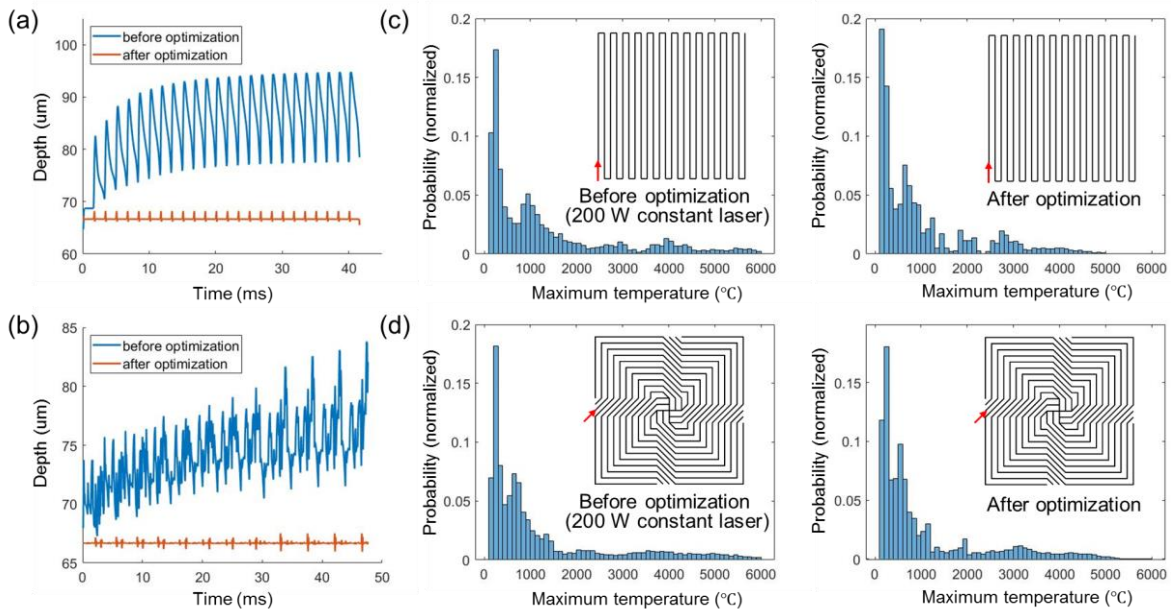

**Figure 6.** Melt pool depth before (constant laser power) and after power optimization obtained for (a) the zigzag scanning pattern and (b) the new spiral scan pattern. The probability distribution of the maximum temperature over the duration of scanning the region with (c) the zigzag pattern and (d) the new spiral scan pattern. The left-side figures in (c,d) correspond to scanning with constant laser power. The right-side figures in (c,d) are for the optimized laser power shown in (a,b).

Figure 4(a) shows the schematic implementation of the model by superposition of spot heat sources. As $\tilde{T}$ is an analytical solution for a semi-infinite space, it does not consider the real boundary conditions defined by the geometry of the printing part. However, in L-PBF, surrounding un-melted powder has very low thermal conductivity, and most heat loss is through the printed part to the build plate. Accordingly, the boundary of the printed part can be assumed to be adiabatic. To correct the solution $\tilde{T}$ by reflecting the boundary conditions, we need to add $\hat{T}$ obtained from a finite element analysis based on equations (5a-d). Here, the boundary condition in (5c) corresponds to the imposition of heat sources at the boundary of the printed part to compensate for the heat flux given from the analytical solution $\tilde{T}$ in the semi-infinite space. Consequently, the superposition solution, $T$, can satisfy the boundary conditions.

We can validate the superposition solution with a case study shown in Figure 4(b). We applied a spiral scan of 2x2 mm at the center of the test region, and then simulated the temperature field along the line A-B in Figure 4(c). The superposition solution, $T$, has a nearly zero slope at the boundary of the printing region, which means the semi-analytical solution can reflect the thermal boundary of the printed part by adding numerical solution $\hat{T}$. Here, $\hat{T}$ was obtained using a Finite Element Method (FEM) implemented in MATLAB. Though a coarse FE-mesh was employed even in the vicinity of the melt pools, the final solution could accurately describe the high gradient solutions near the melt pool.

Now, with this approach, laser power optimization can be implemented on the desired scan pattern. We aim to find the appropriate laser control map that ensures a consistent melt pool depth for the given scan pattern using the semi-analytical method. To optimize the laser power map, $Q^{(I)}$, laser power at each node $I$, was controlled. We applied an iterative local power adjustment scheme based on the calculated melt pool depth, such that

$$Q^{(I)} \leftarrow Q^{(I)} d^*/d^{(I)} \qquad (6)$$

where, $d^*$ and $d^{(I)}$ are the target and calculated melt pool depth, respectively.

In Figure 5, we present a case study of the optimization process for a zigzag scan pattern. In this case, the scan pattern is 2 mm by 2 mm with hatch spacing of 80 μm, and the scan speed is set to 1200 mm/s. The material is Ti-6Al-4V and parameters for the simulation model in (4-6) are taken from (Yang et al. 2018; Yang et al. 2020): $k$ = 42 W/m-K, $\rho$ = 4420 kg/m³, $c_p$ = 990 J/kg-K, and $T_m$ = 1655 °C. The absorptivity, $\tilde{A}$, is set to 0.818, and $T_{base}$ is 100 °C. The laser spot radius, $r$, is 80 μm.

Figure 5(a) shows the zig-zag scan pattern tested for the laser power optimization, and Figure 5(b) corresponds to time histories of laser power and melt pool depth, respectively. The target depth was set to 66.67 μm, and 10 iterations were applied according to (6). Figure 5(b) is the zoomed-in graph for the first two scan vectors. In Figure 5(c), the history for the melt pool depth before optimization (blue color) shows a melt pool depth slightly smaller than the target value at the beginning of the scan. Then, the melt pool depth becomes much larger than the target value after passing through the turning point, which is due to the accumulated heat from the first scan vector. After the optimization routine is applied, the melt pool depth is consistent within 3.2%, and the power periodically modulates between 85 W and 195 W.

In Figure 6, we compare the melt pool depth before (constant power) and after optimization,

for the zig-zag pattern and the proposed spiral pattern, showing that for both patterns power optimization prevents heat accumulation as scanning proceeds over the unit cell. In addition to the melt pool depth, the utility of power optimization can be qualitatively judged by the surface temperature relative to the boiling point (Khairallah et al. 2016). The fraction of the points on the surface that reach the boiling point (3806 °C, (Bartsch et al. 2021)) during scanning was lowered from 12.5 % to 3.55 % for the zigzag scan pattern, and from 11.1 % to 5.68 % for the spiral pattern.

## 4. Evaluation of the spiral scan pattern

Now, the performance of the spiral pattern derived from the predator-prey model is evaluated in comparison to established scan patterns in LPBF, namely the zig-zag pattern and the helix pattern introduced in literature (Bo et al. 2012; Cheng et al. 2016). The primary metric of comparison is the uniformity of the temperature field at the end of the scanning process. A scan pattern that maintains a larger area above a threshold temperature is considered to be desirable for materials that have a specific transition temperature, e.g., a ductile to brittle transition temperature (DBTT) (Vrancken et al. 2020). This could mitigate microcracking issues appearing as a result of the reduced ductility and large residual stresses present below the DBTT, notwithstanding the ability to preheat the specimen during printing (Papadakis et al. 2018; Müller et al. 2019; Caprio et al. 2020). In what follows we apply laser power optimization introduced in Section 3 for all tested scan patterns to reduce the likelihood of defects due to the under/over-melting and evaporation.

Figure 7 shows the considered scan patterns, where two directions, in-out and out-in motions, were evaluated for the rotating spiral patterns and helix patterns. In the case of an out-in motion for the rotating spiral patterns, each scan vector starts from the endpoint located at the boundary of the unit scan pattern. The test scan patterns were drawn on 24 by 24 uniform grid points, and the hatch distance was set to 0.08 mm for all patterns with a spot radius of 80 μm. The baseplate temperature was set to 100 °C.

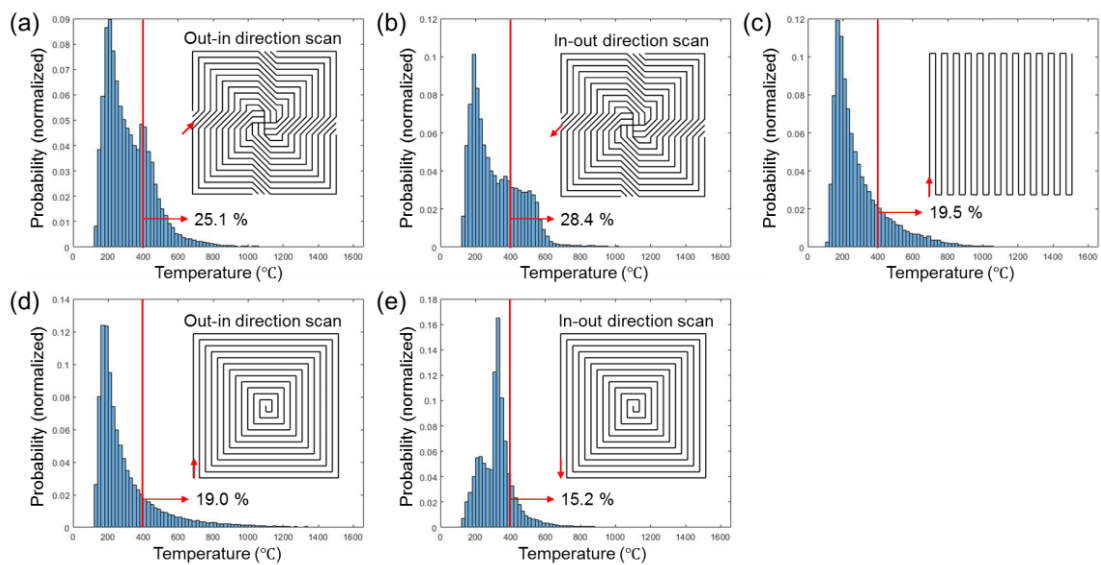

**Figure 7.** Histogram of the temperature fields of the top layer of 150 μm thickness at the end of each scan for (a) rotating spiral pattern (out-in scanning); (b) rotating spiral pattern (in-out

scanning); (c) zig-zag pattern; (d) helix pattern (out-in scanning); and (e) helix pattern (in-out scanning).

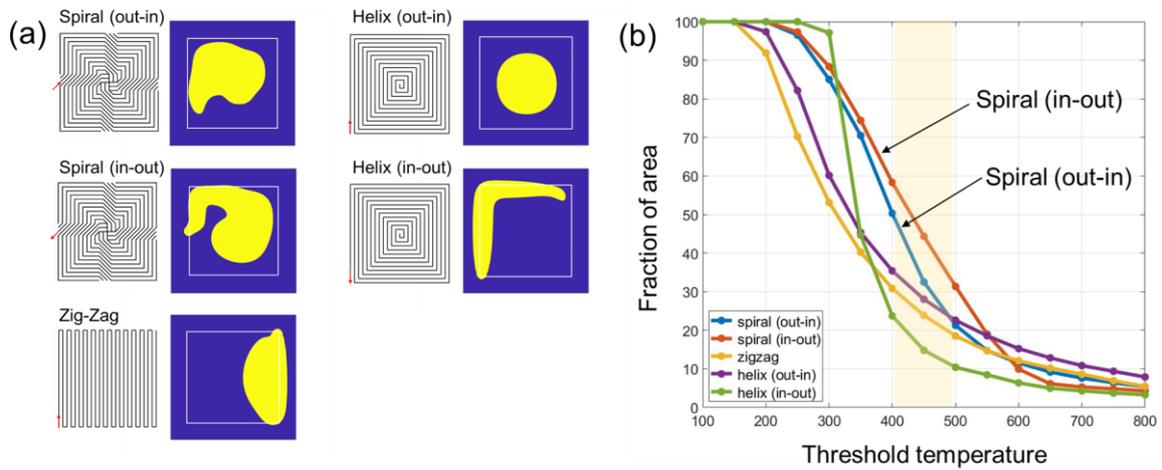

**Figure 8.** (a) Five simulated single-laser scan patterns. The yellow-colored region illustrates the region having a temperature above 400 ℃ at the end of each scanning process. (b) Fraction of the area having a temperature larger than the given threshold temperature at the end of the scanning process for the five tested patterns.

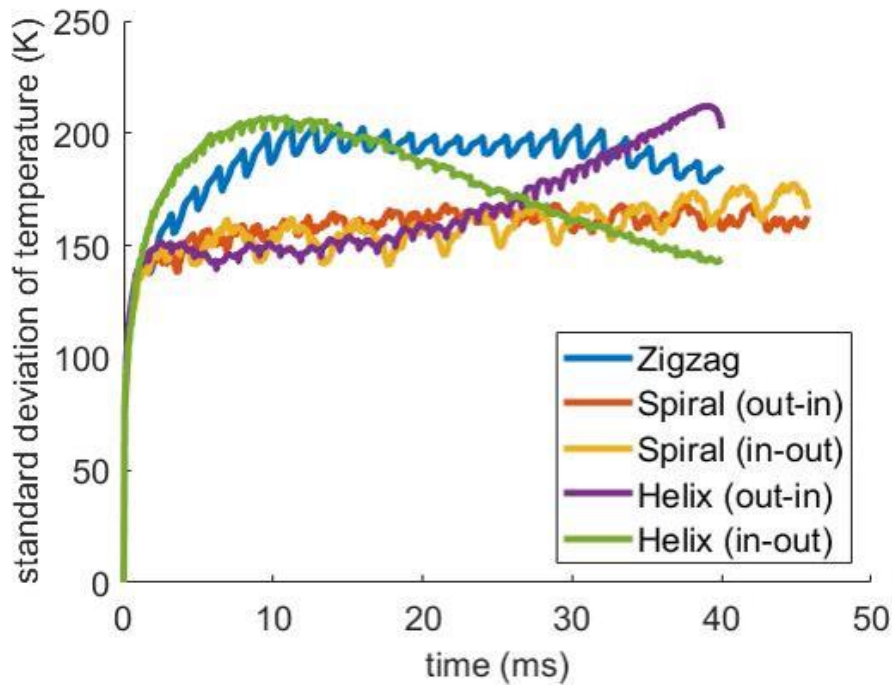

**Figure 9.** Time history of the standard deviation of temperature fields for each scan pattern, versus the time during scanning, calculated over the full area and 150 μm thickness beneath the top surface.

Figure 7 shows the probability distribution of the temperature for points in the scanning region, immediately following the completion of each scan. Here, we calculated the temperature at the grid points evenly distributed in the top surface of 150 μm thickness corresponding to three times the assumed printed layer height (50 μm). Considering that the printing process generally remelts a part of one or two layers below the current printing layer, we established the three layers as relevant for comparison of the thermal fields. Compared with other patterns, the probability of having a temperature above the critical value, e.g., DBTT set to 400 °C, is higher with the spiral patterns shown in Figure 7(a,b). Similarly, we can also calculate the area fraction of the scanning region maintaining temperature above the specific value after finishing the scanning process. Figure 8(a) shows the area with a temperature higher than 400 °C, and Figure 8(b) represents the fraction of the scanning area having a temperature above the given threshold temperature (called the heated-area ratio), e.g., 400 °C. The rotating spiral patterns have a heated-area ratio larger than that of the other scan patterns.

The time history of the temperature field during the scanning process is also instrumental to material quality resulting from LPBF. Ideally, uniform heating and cooling will mitigate variation and defects caused by inhomogeneous temperature fields, e.g., morphologies of the microstructures are influenced by the spatial temperature gradient around the melt pool and local cooling rates. Figure 9 represents the standard deviation of the temperature fields calculated at each time step of the scanning process for the five different scan patterns. The helix pattern scanned by in-out direction shows the smallest standard deviation at the end of the scanning process, but the proposed rotating spiral patterns show consistent standard deviation over the whole scanning process. For the rotating spiral patterns, the overall level of the variation is also smaller than that of other patterns, which means that heating and cooling by the proposed scan pattern is consistent and uniform compared with other patterns. The consistent spatial uniformity is due to the spiral trajectory and the order of the scan vectors, where the pattern is completed by sequentially "filling in" the pattern while maintaining a constant distance between sequentially scanned tracks.

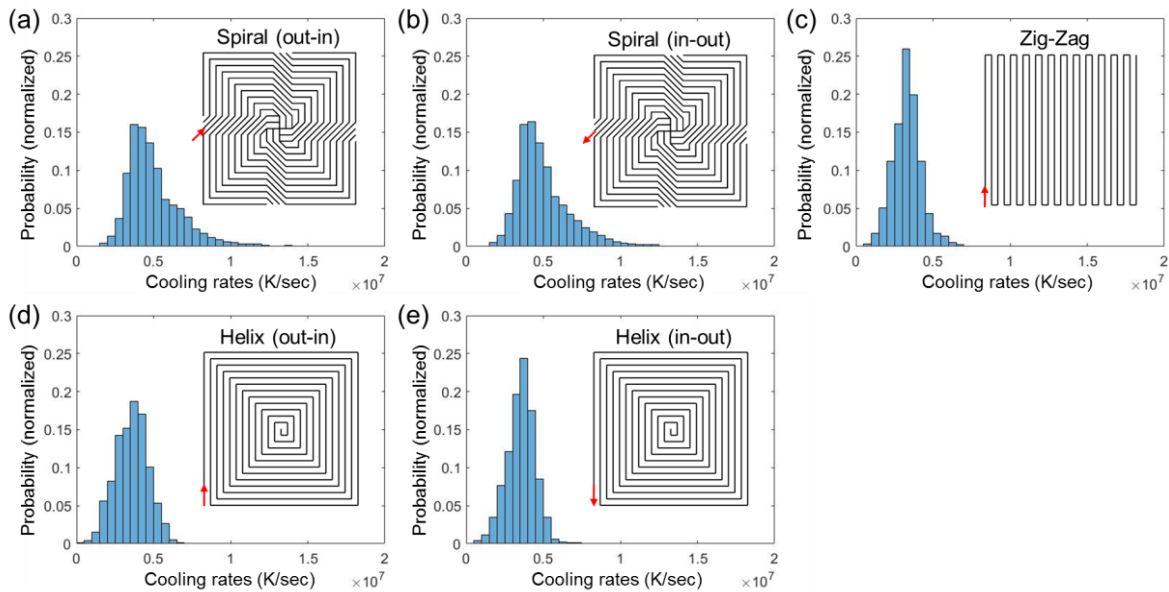

**Figure 10.** Histogram of the cooling rates measured over the whole scanning process: (a) Rotating spiral pattern (out-in scanning); (b) Rotating spiral pattern (in-out scanning); (c) Zig-zag pattern; (d) Helix pattern (out-in scanning); (e) Helix pattern (in-out scanning). The cooling rates were calculated at each grid point at the time when the local temperature passes through the melting point.

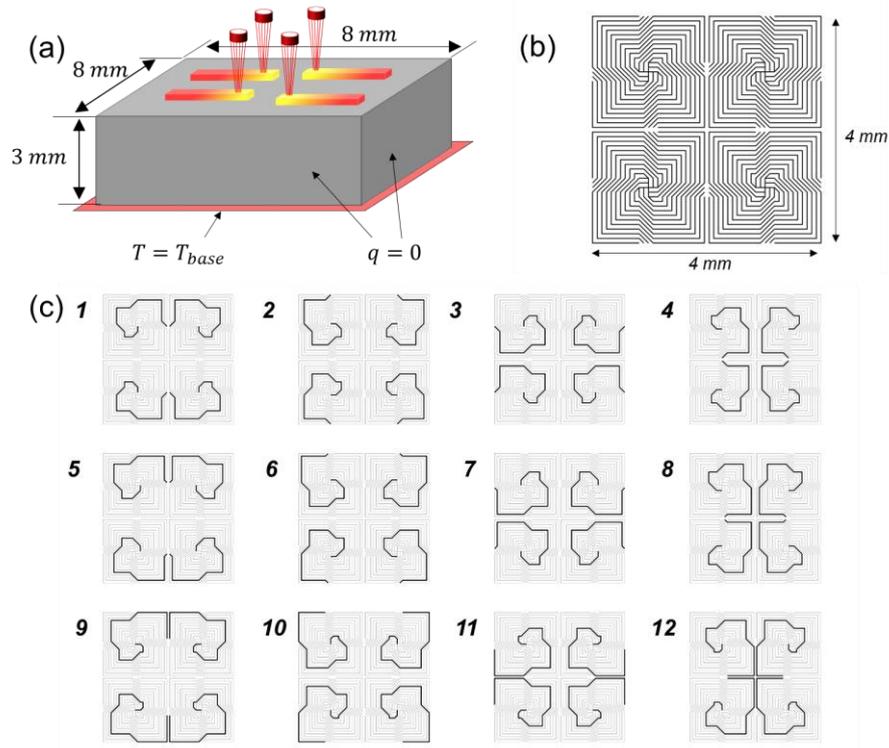

**Figure 11.** (a) Scan region and boundary conditions applied to the quad laser case study. (b) A proposed pattern for simultaneous scanning of four lasers. (c) Twelve sets of scan vectors composing the quad laser scan pattern (all scan vectors can be drawn outside-inside or vice versa).

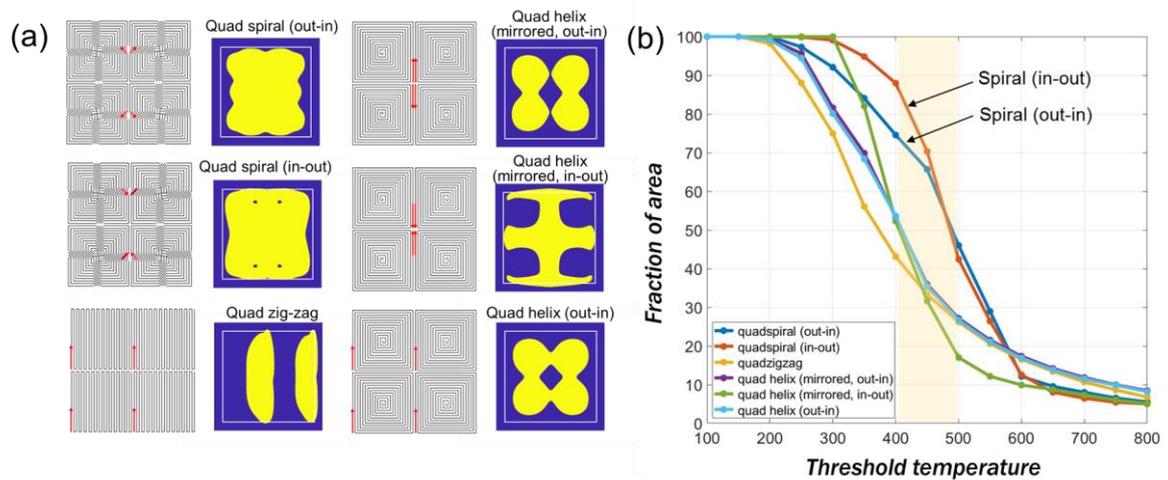

**Figure 12.** Evaluation of quad-laser scan patterns. (a) The yellow-colored region illustrates the region having a temperature above 400 °C at the end of each scanning process. (b) Fraction of the area having a temperature above the given threshold temperature at the end of the scanning process for the six tested patterns.

In Figure 10, we plot the distribution of simulated cooling rates, measured as of when the temperature at each location on the scan pattern passes through the melting point. Because sequential scan vectors are not adjacent, the spiral pattern gives a higher maximum cooling rate than other scan patterns. The result is well-matched with the nature of prey behavior moving to avoid predators. In the Appendix A (Figure A.1) we show temperature gradients plotted versus solidification rates, calculated from the local cooling rates divided by the temperature gradient. The distribution of cooling rates and solidification rates is comparable among all patterns, because the longer-time cooling behavior is determined primarily by the surroundings, e.g., the part geometry. Therefore, the order at which islands are visited by the scanning algorithm, or the dimensions of the part (e.g., thin walls, islands) should be considered in scan parameter assignment to tailor the thermal history over short- and long times.

Last, we demonstrate the extension of the spiral pattern to the case of multiple independently scanned lasers, which are more commonplace in commercial LPBF equipment, and desirable for achieving higher printing throughput. The simulation result of Figure 2 showed multiple rotating spirals with symmetry, which inspired synthesizing a quad laser scan pattern involving four rotating spirals shown in Figure 11. Practically, independently scanned spirals may address areas at any location of the build area; here, we consider the case if each simulated laser spot addresses an adjacent scan area. To compare other patterns under a quad laser system assumption, the helix and zig-zag scan patterns were extended to the scan patterns for the quad laser system in the similar fashion, as shown in Figure 12. Though the testing patterns did not consider practical constraints – e.g., preventing lasers from approaching close to each other, the patterns are enough to see the effectiveness of the spiral pattern inspired by the predator and prey behavior. Since heat mainly diffuses at the boundary of the unit cell, in-out direction scanning can be advantageous to manage the temperature fields uniformly. For example, the highest value (87.96 % above 400 °C) was obtained by a quad spiral pattern with an in-out scanning direction. Figure 13 shows the temperature field that resulted from the quad spiral pattern. It represents a similar configuration as a quarter of the predator map shown in Figure 2 (see the predator map at t=368.0 ms), even though the applied scan pattern was adapted for the practical implementation by a raster scanning process. Thus, we can accept the quad spiral pattern as well-designed to mimic the predator-prey behavior.

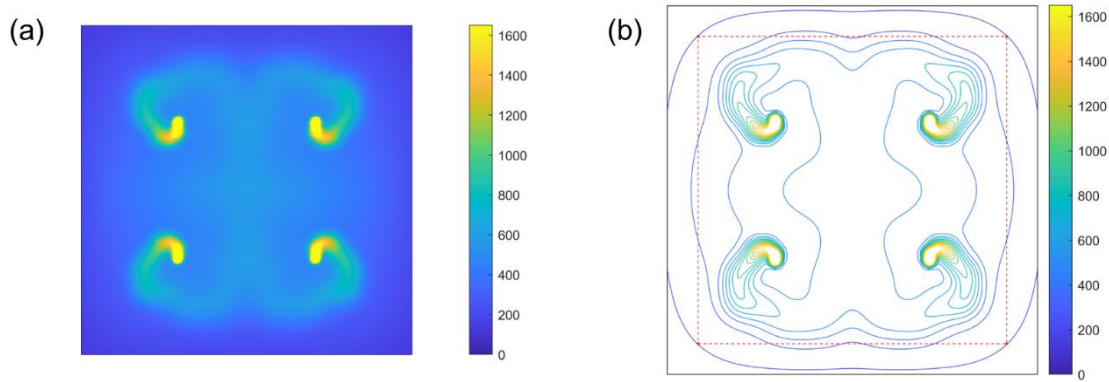

**Figure 13.** (a) Temperature fields and (b) contour map calculated at the end of the scanning process for the quad rotating spiral pattern (out-in scanning).

## 5. Conclusion

A new scan pattern, called the rotating spiral pattern, was proposed based on the similarity between the predator-prey model and the LPBF process. After deriving the nature-inspired pattern, laser power map was optimized based on the semi-analytical simulation to avoid defects due to the inconsistent melt pool depth and too high a temperature reaching the boiling point. Compared with other patterns previously studied, the rotating spiral pattern showed more uniform and consistent heating and cooling during the process, and this property was observed from the time history graph for the standard deviation of the temperature fields. Additionally, simulations of the temperature distribution under spiral scanning predicts improved heat retention and thermal uniformity. From that, we can expect the rotating spiral pattern will help address thermally-induced problems due to non-uniform heating and cooling in the LPBF process in concert with other process control schemes, such as preheating of the build platform. Though the new scan pattern was designed without consideration of arbitrary boundary shapes in this research, the proposed design method can also be applied to find the solution for the specific geometry, e.g., overhangs or bridges, challenging to print by the LPBF process without enough supporting structures. To this end, we can apply the predator-prey model to derive the scan pattern suitable for the given geometry by simulating the model with proper boundary conditions. The framework proposed in this research does not require an iterative procedure for path-planning, so it can be a computationally efficient tool for deriving scan patterns of each layer of the part, regardless of its contour shape. The proposed framework could be extended to control other additive manufacturing processes using lights or heat sources, including electron beam melting or area-wide LPBF using modulated laser arrays (Roehling et al. 2021), and it will also be applicable for other limited resource management problems with the spatiotemporal system showing a time-decaying pattern. Experimental validation of the scan patterns, including by in situ thermal imaging, material quality (e.g., density, microstructure), and dimensional quality (e.g., geometric distortion due to residual stresses) is also necessary in future work.

## Conflicts of interest

The authors declare no conflicts of interest.

## Acknowledgements

We acknowledge funding from the Ford-MIT alliance, the MIT-Portugal Program (MPP2030), the Advanced Research Projects-Energy (ARPA-E, award #DE-AR0001434), and the MIT Center for Additive and Digital Advanced Production Technologies (APT).

## Notes on contributors

**Suh In Kim** received BS (2011) and MS-PhD combined (2017) degrees from Seoul National University in Korea. He was a research engineer at Hyundai Motor Company, and he is currently a postdoctoral associate at Massachusetts Institute of Technology in the USA. His research interests lie at the intersection of automatic design and digital fabrication. He is doing research for the product design and process optimization for additive manufacturing.

**A. John Hart** received his BS (2000) from the University of Michigan, and MS (2002), and PhD (2006) degrees from the Massachusetts Institute of Technology. At MIT, he is Professor of Mechanical Engineering, Director of the Center for Additive and Digital Advanced Production Technologies (APT), and Director of the Laboratory for Manufacturing and Productivity (LMP). His research focuses on materials, processes, and computational techniques for additive manufacturing; nanostructured materials and devices; and precision machine design. Hart is a co-founder of startup companies Desktop Metal and VulcanForms, and a Board Member of Carpenter Technology Corporation.

# Appendix A

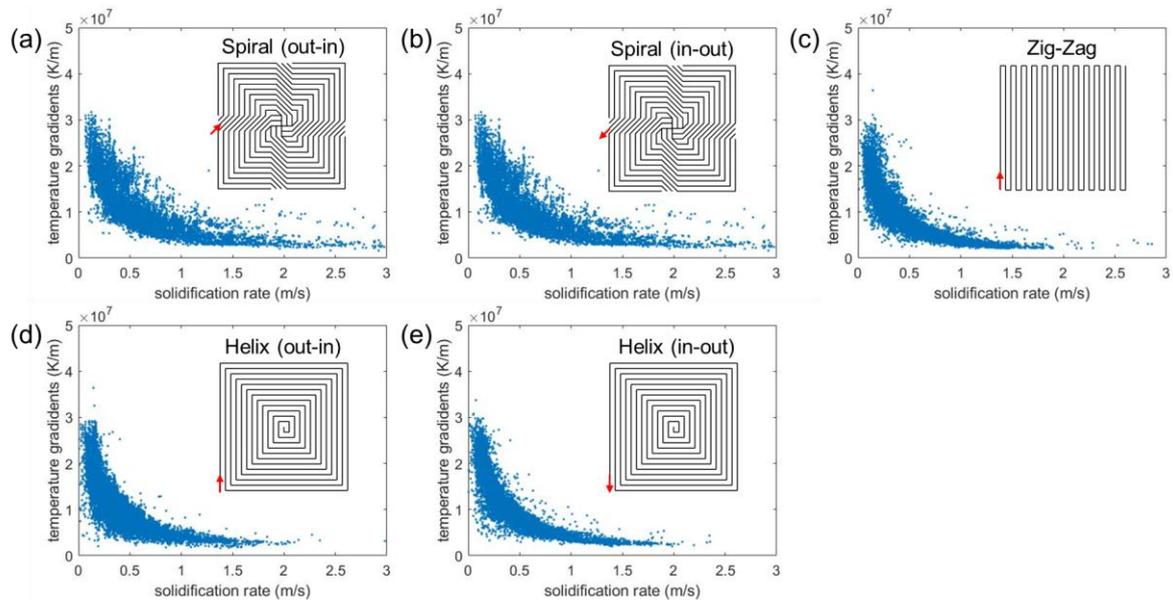

**Figure A.1.** Graphs of the solidification rate and temperature gradients of each scan pattern. (a) Rotating spiral pattern (out-in scanning) (b) Rotating spiral pattern (in-out scanning) (c) Zig-zag pattern (d) Helix pattern (out-in scanning) (e) Helix pattern (in-out scanning).